\documentclass[twocolumn,showpacs,preprintnumbers,amsmath,amssymb]{revtex4}

\usepackage{graphicx}
\usepackage{dcolumn}
\usepackage{bm}
\usepackage{epstopdf}

\begin{document}

\preprint{}

\title{Few-femtosecond Electron Beam with THz-frequency Wakefield-driven Compression}

\author{Lingrong Zhao$^{1,2}$, Tao Jiang$^{1,2}$, Chao Lu$^{1,2}$, Rui Wang$^{1,2}$, Zhe Wang$^{1,2}$, Pengfei Zhu$^{1,2}$, Yanchao Shi$^{3}$, Wei Song$^{3}$, Xiaoxin Zhu$^{3}$, Chunguang Jing$^{4}$, Sergey Antipov$^{4}$, Dao Xiang$^{1,2,5*}$ and Jie Zhang$^{1,2\dagger}$}

\affiliation{%
$^1$ Key Laboratory for Laser Plasmas (Ministry of Education), School of Physics and Astronomy, Shanghai Jiao Tong University, Shanghai 200240, China \\
$^2$  Collaborative Innovation Center of IFSA (CICIFSA), Shanghai Jiao Tong University, Shanghai 200240, China \\
$^3$ Science and Technology on High Power Microwave Laboratory, Northwest Institute of Nuclear Technology, Xi'an, Shanxi 710024, China\\
$^4$ Euclid Techlabs LLC, Bolingbrook, Illinois 60440, USA \\
$^5$  Tsung-Dao Lee Institute, Shanghai 200240, China \\
}
\date{\today}

\begin{abstract}
We propose and demonstrate a novel method to produce few-femtosecond electron beam with relatively low timing jitter. In this method a relativistic electron beam is compressed from about 150 fs (rms) to about 7 fs (rms, upper limit) with the wakefield at THz frequency produced by a leading drive beam in a dielectric tube. By imprinting the energy chirp in a passive way, we demonstrate through laser-driven THz streaking technique that no additional timing jitter with respect to an external laser is introduced in this bunch compression process, a prominent advantage over the conventional method using radio-frequency bunchers. We expect that this passive bunching technique may enable new opportunities in many ultrashort-beam based advanced applications such as ultrafast electron diffraction and plasma wakefield acceleration.  
\end{abstract}

\maketitle
Ultrashort electron beams are of fundamental interest in accelerator physics and ultrafast science communities. In x-ray free-electron lasers (FELs \cite{LCLS, SACLA, PAL}), ultrashort electron beams with high peak current and low emittance are used to produce coherent and intense x-rays that have already opened new opportunities in many areas of science \cite{LCLS5years}. In keV and MeV ultrafast electron diffraction (UED \cite{UED1, UED2, UED3, UCLA, THU, OSAKA, SJTU, BNL, PKU, SLAC, DESY}), ultrashort electron beams are used to probe the atomic structure changes in many non-equilibrium processes. In beam-driven plasma wakefield accelerator (PWFA \cite{ED, FACET}), ultrashort electron beams are essential for exciting the high-gradient plasma wakefield. Ultrashort electron beams are also important for producing intense terahertz (THz) pulses \cite{SDL, LCLS-THz}. 

For an electron beam that contains millions of electrons, Coulomb repulsion force tends to broaden the pulse width and in general ultrashort electron beams are obtained with bunch compression \cite{JAP, CPRL, CUCLA}. The process requires first a mechanism to imprint energy chirp (correlation between a particle's energy and its longitudinal position) in the beam longitudinal phase space and then sending the beam through a dispersive element such that the longitudinal displacement of the electrons is changed in a controlled way to yield a beam with shorter pulse width. The required energy chirp is typically established by accelerating the beam off-crest in radio-frequency (rf) cavities and the widely used dispersive elements are magnetic chicanes for ultra-relativistic electron beam and drifts for near-relativistic and sub-relativistic beams. 

    \begin{figure*}[t]
    \includegraphics[width = 0.9\textwidth]{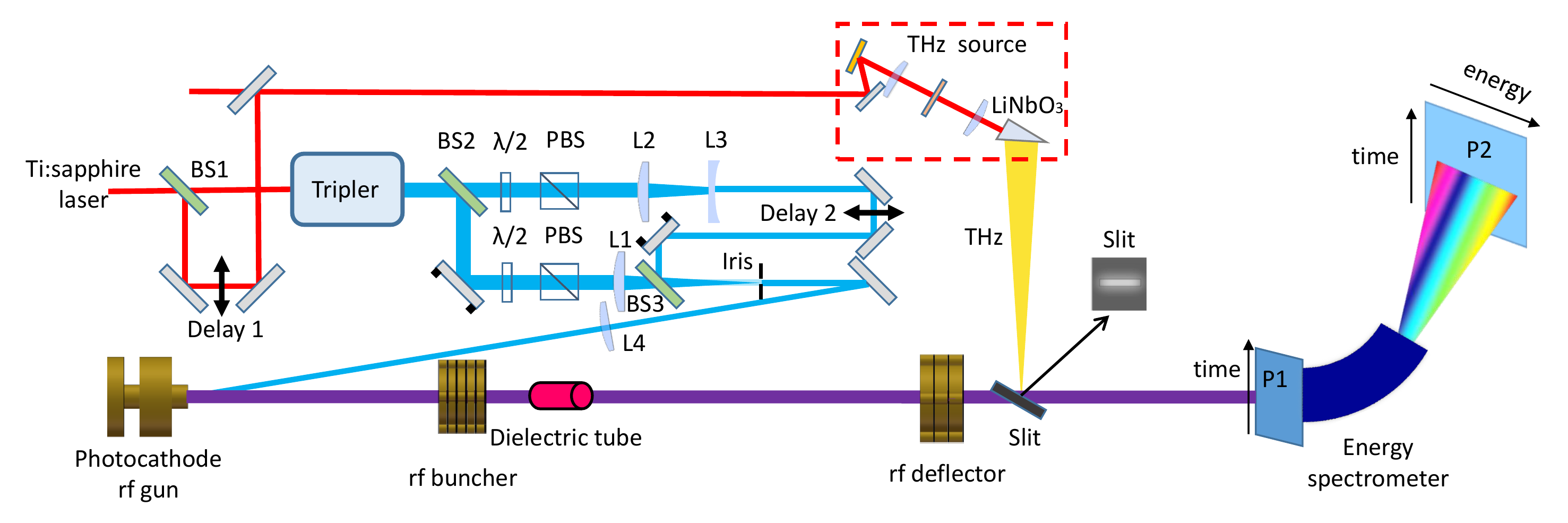}
            \caption{Wakefield-driven relativistic electron beam compression experiment setup. The 800 nm laser is split into three pulses with two pulses used for producing electron beams and the third pulse for producing THz radiation. The trailing witness beam is compressed by the wakefield generated by the leading drive beam in a dielectric tube. The temporal profile of the witness beam is measured with an rf deflector and its arrival time is measured with THz streaking in a slit. 
    \label{Fig.1}}
    \end{figure*}

The main drawback of this active bunching technique is that the phase jitter in the rf cavity will be converted into timing jitter after compression \cite{RFC1, RFC2}. This is because the rf phase jitter leads to variation of the beam centroid energy, which is further translated to variation of time-of-flight after passing through a dispersive element. In this Letter, we demonstrate a passive bunching technique where an electron beam is compressed from about 150 fs (rms) to about 7 fs (rms, upper limit) with the wakefield produced by a leading drive beam in a dielectric tube. Furthermore, with laser-driven THz streaking technique \cite{TS1, TS2}, it is also demonstrated that this passive compression scheme introduces negligible additional timing jitter with respect to an external laser due to the fact that the wakefield is naturally synchronized with the electron beams. This passive bunching technique is cost effective, easy to implement, and should find wide applications in many ultrashort-beam based advanced applications. 

The schematic layout of the experiment is shown in Fig.~1. A Ti:sapphire laser at 800 nm is first split into two pulses with a 50\%-50\% beam splitter (BS1). One pulse is ($\sim$2 mJ) used to produce THz radiation through optical rectification in LiNbO$_3$ crystal with the tilted-pulse-front-pumping scheme \cite{TPFP}. The other pulse is frequency tripled to produce electron beams in a 1.5 cell S-band (2856 MHz) photocathode rf gun. The UV laser is further split into two parts with a 4\%-96\% beam splitter (BS2) with the main pulse used to produce the leading drive beam and the remaining part used to produce the trailing witness beam. The two UV pulses are combined with a 50\%-50\% beam splitter (BS3). Such a setup allows us to vary the energy, transverse size and timing of the three laser pulses independently. The drive beam produces a wakefield in a dielectric tube which imprints negative energy chirp (i.e. with the bunch head having a lower energy than the bunch tail) in the witness beam that allows bunch compression after a drift. Such wakefields have also been used for generation of THz pulses \cite{THz2012, THz2013, THz2017}, beam acceleration \cite{TA, NC} and manipulation \cite{Bane12, EM1, Emma, EM2, Deng, SJTUEM, NPLCLS, PassiveD, PL}. The electron beam temporal profile is measured with a 5-cell c-band (5712 MHz) deflecting cavity at a screen (P1). Together with the energy spectrometer, the beam longitudinal phase space can also be measured at a screen (P2). The arrival time jitter of the compressed beam is measured through THz streaking technique \cite{TS1, TS2} just downstream of the rf deflector. 

    \begin{figure}[b]
    \includegraphics[width = 0.49\textwidth]{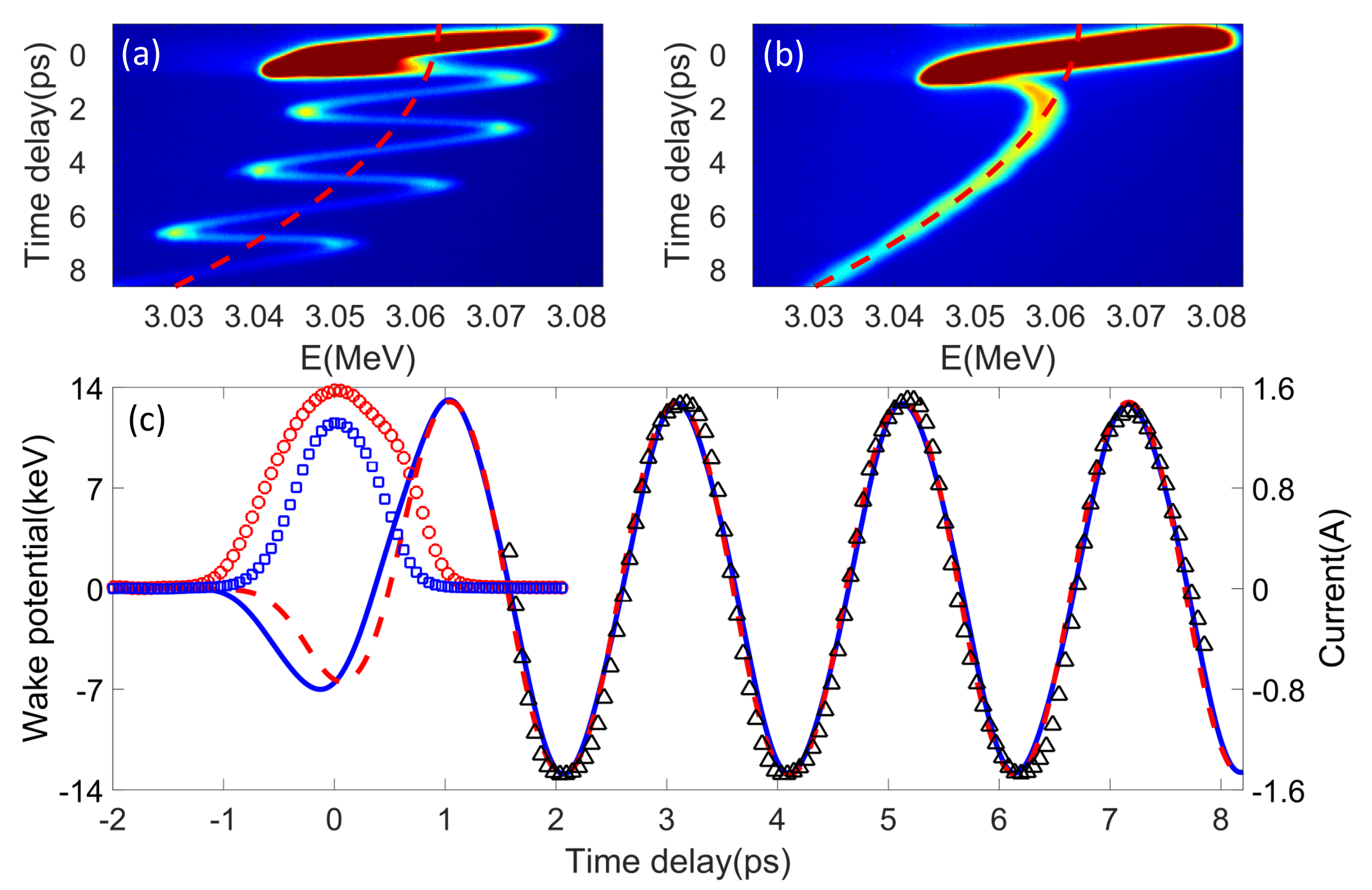}
            \caption{Electron beam time-energy map with (a) and without (b) the dielectric tube. The red dashed line shows the curvature of the rf field and the saturated region represents the drive beam with high intensity. (c) Measured longitudinal wake potential (black triangles) and the calculation (red dashed line) using the measured beam distribution at the deflector (red circles). For comparison, the retrieved leading beam distribution at the dielectric tube (blue squares) and its corresponding wake potential (blue solid line) are also shown.   
    \label{Fig.2}}
    \end{figure}

In this experiment, a 5 cm long quartz capillary tube with inner diameter of 400 $\mu$m is used as the wakefield structure. The charge of the drive beam and witness beam is measured to be about 1.0 pC and 10 fC, respectively. With the gun solenoid used to optimize the beam size at the dielectric tube, both the drive beam and witness beam can pass through the tube with negligible loss. The longitudinal wake potential produced by the drive beam is mapped through measurement of the energy change of the witness beam as a function of separation of the two bunches. In this experiment, single-shot measurements of the beam distribution on screen P2 with both the deflector and energy spectrometer on for various separation of the two bunches is first measured and then superimposed. This is equivalent to the measurement with a long witness beam extending over several periods of the wakefield. After converting the time at the deflector back to that at the dielectric tube using the momentum compaction of the drift, the corresponding time-energy map with and without the dielectric tube is obtained and shown in Fig.~2a and Fig.~2b, respectively. Without the tube, the time-energy map follows the curvature of the rf field (dashed line in Fig.~2b) when the witness beam is far away from the drive beam. Considerable deviation from the rf curvature is observed when the witness beam is close to the drive beam, which is due to space charge effect \cite{CD}. Due to similar effect, the longitudinal phase space of the drive beam is dominated by positive chirp (saturated region in Fig.~2a and 2b). With the dielectric tube inserted, considerable energy modulation from the wakefield is imprinted in the witness beam time-energy map (Fig.~2a).

From Fig.~2a and Fig.~2b, the wake potential is obtained and shown in Fig.~2c (black triangles). The wake potential can be accurately approximated by a single mode with central frequency around 0.5 THz. To compare the measured wake potential with the theory, the simulated point charge wake is convoluted with the measured leading beam distribution (red circle) and the calculated wake potential fits the data quite well when the beam charge is taken to be 2.1 pC. It should be pointed out that when calculating the wakefield, the leading beam distribution at the dielectric tube should be used. However, in our experiment the leading beam distribution is measured at the deflector which is 1.1 m downstream of the dielectric tube. As a first approximation we back propagate the beam distribution from the deflector to the dielectric tube (space charge effect is not included), and the corresponding leading beam distribution is shown with blue squares in Fig.~2c. Using the retrieved beam distribution at the dielectric tube, the calculated wake potential can also fit the data quite well if the beam charge is taken to be 1.2 pC. The main reason for the different charges that provide the best fit is that the shorter bunch has stronger frequency component at 0.5 THz. It is worth mentioning that when space charge effect is taken into account, the bunch length should be shorter than that obtained with simple back tracking and thus the required charge to produce the measured wake potential should be lower than 1.2 pC, consistent with the measured beam charge in this experiment. 

In conventional bunching technique, the electron beam is typically sent through an rf cavity at the zero-crossing phase and the normalized energy chirp may be written as $h=-2\pi V/E\lambda$, where $V$ is the voltage of the cavity, $E$ is the beam energy and $\lambda$ is the wavelength of the field. With the space charge effect neglected, analysis shows that full compression is achieved when the beam is further sent through a dispersive element with momentum compaction $R_{56}=-1/h$ (see, e.g. \cite{RMP}). In our experiment ($R_{56}\approx3.07$ cm and $E\approx$3.06 MeV), a C-band buncher cavity with about 1 MV voltage is needed to compress the beam, which requires an expensive rf station. Alternatively, with a wakefield having 100 times higher frequency, the required voltage is reduced to about 10 kV, which may be readily produced by a beam with $\sim$pC charge, as shown in Fig.~2.

To compress the witness beam, a delay stage (Delay 2 in Fig.~1) is used to adjust the timing of the witness beam to make it ride at the negative zero-crossing phase (e.g. at a time delay of 2.6 ps in Fig.~2c). The strength of the wakefield is varied by changing the charge of the drive beam and the corresponding longitudinal phase space of the witness beam measured at screen P2 is shown in Fig.~3a-d. Without the drive beam, the witness beam has slightly positive chirp in the phase space from space charge force as shown in Fig.~3a. As the charge of the drive beam is gradually increased, the energy chirp of the witness beam is reversed and the beam is longitudinally compressed (Fig.~3b). The witness beam is fully compressed in Fig.~3c and further increasing the drive beam charge leads to over-compression and the bunch length starts to increase (Fig.~3d). In this measurement, the voltage of the rf deflector is set at about 0.2 MV, sufficient to resolve the longitudinal phase space while still keeping the energy spread growth in the deflector (see, e.g. \cite{echo7, RH, BDD}) at a small level. It should be pointed out that while the leading beam temporal profile also changes with the beam charge, the position of the zero-crossings of the wakefield (Fig.~2c) is insensitive to the temporal profile of the leading beam, which makes it possible to tune the energy chirp for optimal compression without changing the centroid energy of the witness beam.

    \begin{figure}[b]
    \includegraphics[width = 0.49\textwidth]{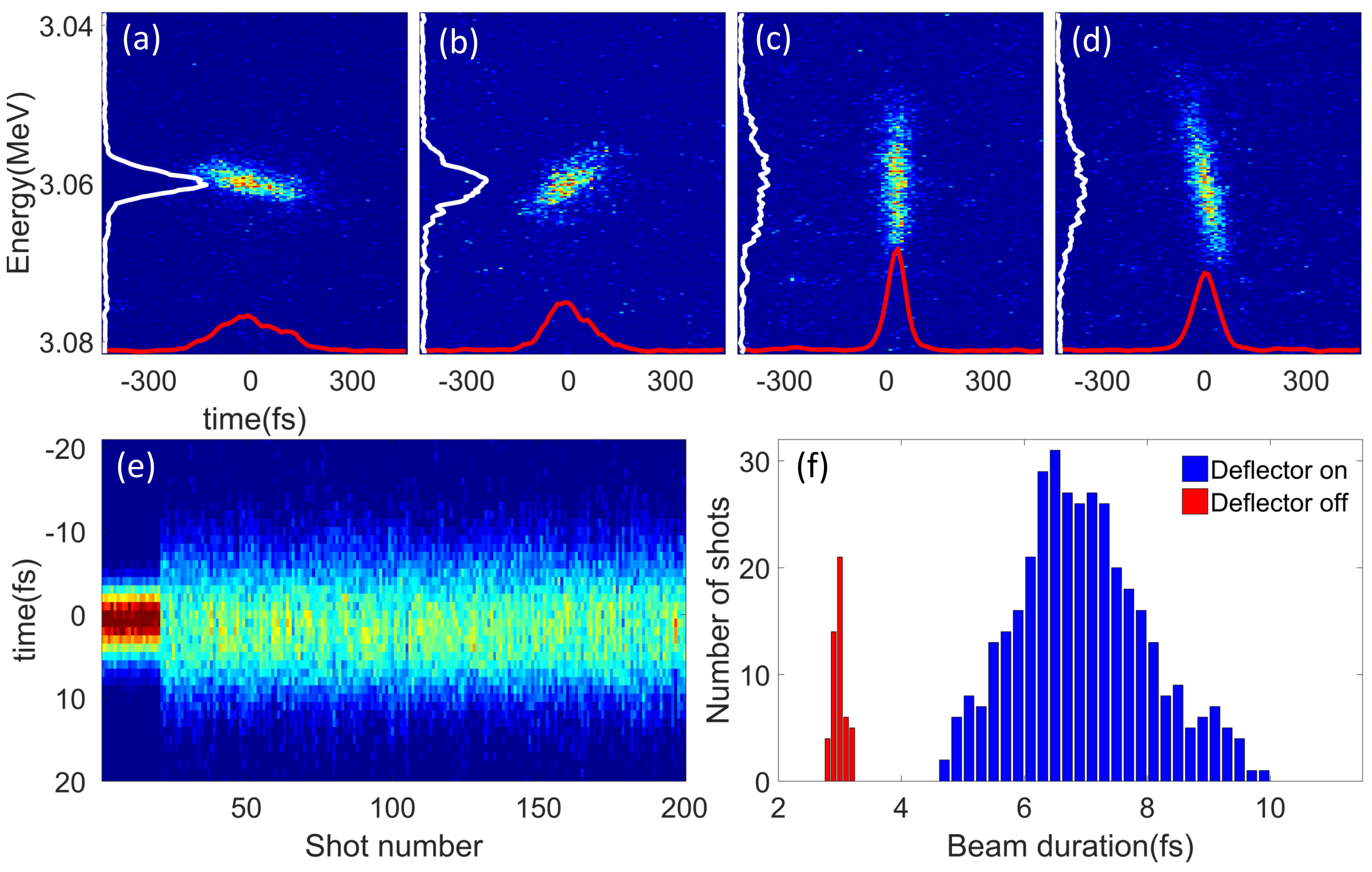}
            \caption{Longitudinal phase space (bunch head to the left) of the witness beam when the charge of the drive beam is changed from 0 (a) to 0.6 pC (b), 0.9 pC (c), and 1.3 pC (d). The corresponding temporal profile (red line) and energy distribution (white line) are also shown. (e) 200 consecutive measurements of the witness beam temporal profile with rf deflector off (the first 20 shots) and on (the rest 180 shots). (f) Distribution of the raw bunch length collected over 50 consecutive shots with deflector off and 400 consecutive shots with deflector on.
    \label{Fig.3}}
    \end{figure}

To measure the fully compressed beam accurately, the voltage of the deflector is increased to about 1.8 MV which yields a calibration coefficient of about 25 fs/mm on screen P1. With a 20 microns slit inserted before the deflector, the vertical beam size on screen P1 is about 0.11 mm (rms) when the deflector is off, corresponding to 2.7 fs (rms) resolution. Such high temporal resolution allows us to unambiguously measure the strongly compressed beam with pulse width well below 10 fs. Under full compression condition, 200 consecutive measurements of the raw beam profile (with vertical axis converted into time) with rf deflector off and on are shown in Fig.~3e. 50 consecutive shots of the raw beam profile with deflector off and 400 consecutive shots with deflector on are fitted with Gaussian functions and the distribution of the fitted beam duration (rms) is shown in Fig.~3f where one can see that the average bunch length is about 7 fs and the shortest bunch length (3 shots out of 400) is about 4.6 fs (rms). With the contribution from beam intrinsic divergence being 2.7 fs and that from high order effects estimated to be about 2.5 fs \cite{BDD}, the deconvoluted bunch length is found to be about 2.8 fs, indicating that an unprecedented short bunch might be produced. 

With the energy chirp imprinted in a passive way, the energy stability of the witness beam is measured to be about $2.4\times10^{-4}$ (rms) with and without the drive beam, indicating that no additional timing jitter is introduced in this passive bunching technique. In contrast, the energy jitter increases to about $1.5\times10^{-3}$ (rms) when the rf buncher is used to compress the witness beam, implying that the beam is compressed at the cost of considerably increased timing jitter.

To directly measure the beam arrival time jitter, a single-cycle THz pulse produced by the 800 nm laser in a LiNbO$_3$ crystal is focused to a $250\times15$ microns narrow slit to streak the witness beam \cite{TS2}, similar to a THz deflecting cavity. The THz pulse is linearly polarized along the slit's short axis, allowing the streaking strength to be enhanced (see, e.g. \cite{TE}). The THz field gives the electron beam a time-dependent angular kick which allows us to map the electron beam time information into spatial distribution at screen P1. After the spatial and temporal overlap between the THz beam and the witness beam (the drive beam is off) is achieved in the slit, the timing of the THz beam is varied (Delay 1 in Fig.~1) and the measured streaking deflectogram is shown in Fig.~4a (the oscillation is due to transmission resonance in the slit). 

    \begin{figure}[b]
    \includegraphics[width = 0.49\textwidth]{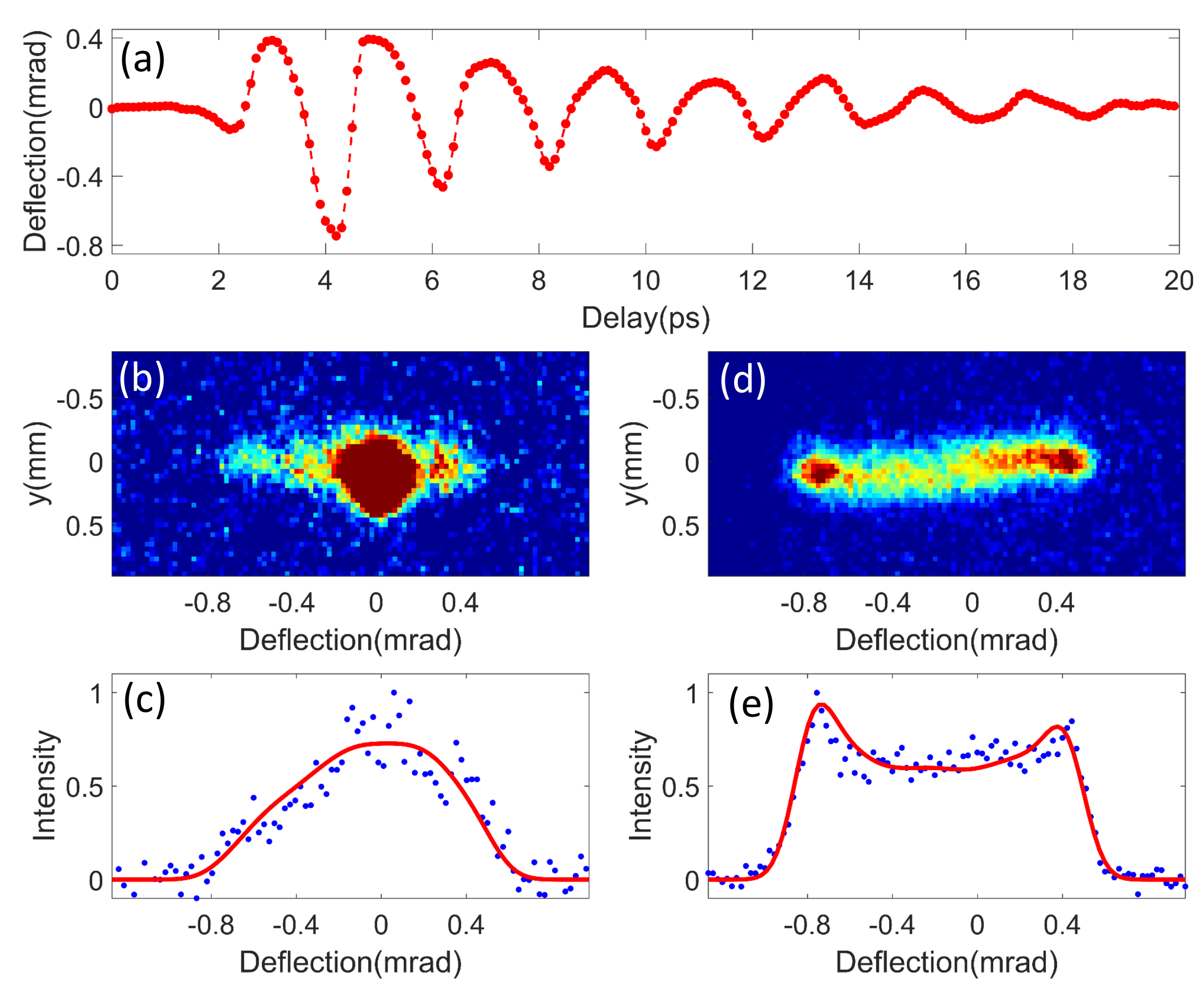}
            \caption{Time-stamping of the compressed beam. (a) THz streaking deflectogram measured as a function of time delay (in 100 fs steps) between the electron beam and THz pulse. Raw distribution of the streaked beam with passive bunching technique (b) and fitting to the timing data (c). Raw distribution of the streaked beam with active bunching technique (d) and fitting to the timing data (e).  
    \label{Fig.4}}
    \end{figure}

Similar to the measurement with an rf deflector, the electron beam should overlap with the THz pulse near the zero-crossing of the deflectogram for measurement of beam timing jitter. The maximal streaking ramp (around t=4.6 ps region in Fig.~4a) is found to be 4.4 $\mu$rad/fs. With the fluctuation of the beam centroid divergence measured to be 10 $\mu$rad, the accuracy of the arrival time measurement is found to be about 2.3 fs. For wakefield-driven compression, the measured streaked beam distribution with integration over 200 shots is shown in Fig.~4b. Note, the drive beam arrives at the slit earlier than the THz pulse, so it is not streaked (see the bright spot in Fig.~4b). After subtracting the background measured with only the drive beam, the distribution of the streaked witness beam is shown in Fig.~4c (blue dots) and a fit (red line) to the data yields a timing jitter of about 60 fs (rms) for the passively compressed beam. This value is consistent with the theoretical jitter taking into account the rf amplitude stability (about $2.4\times10^{-4}$) and the momentum compaction from the gun to the THz slit (about 7 cm). For comparison, in a separate experiment the witness beam is compressed with the rf buncher and the corresponding streaked beam distribution integrated over 200 shots is shown in Fig.~4d. The streaked beam has a double-horn distribution, indicating that the arrival time extends over half of the period of the deflectogram. Assuming the timing jitter has a Gaussian distribution and the best fit to the data (red line in Fig.~4e) yields a timing jitter of about 170 fs (rms) for the actively compressed beam, much larger than that achieved with passive bunching technique. 

It should be mentioned that the THz streaking technique measures the electron beam arrival time with respect to an external laser, which is of interest for pump-probe applications. While in our experiment this value is about 60 fs (with improved rf stability and reduced drift distance, this value can be made much smaller than 50 fs), the timing jitter between the drive beam and the compressed witness beam is negligibly small, which may enable new applications in PWFA. For a beam-driven PWFA, external injection of the witness beam allows precise control of the beam charge and quality. However, the timing jitter between the drive beam and externally injected witness beam limits the energy stability of the witness beam after acceleration. Also the pulse width of the witness beam needs to be much smaller than the wavelength of the plasma field to reduce beam energy spread. The aforementioned challenges may both be overcome with the passive bunching technique. In this case, as long as the injection timing jitter of the witness beam is smaller than half of the wakefield wavelength in the dielectric tube, the witness beam duration and timing jitter will both be compressed after a chicane. Then the compressed witness beam may be further accelerated in the plasma wakefield produced by the tightly synchronized leading beam. This configuration holds great potential for producing stable high quality beams in PWFA. 

Finally, we discuss applications of this passive compression technique in keV UED where it may allow one to produce a shorter beam than that with rf buncher. This is because for optimal compression a linear velocity chirp is needed while the rf buncher only imprints a linear energy chirp in the beam phase space. Note, the kinetic energy of the electron beam at low energy is proportional to the square of the beam velocity, so considerable nonlinearity is present in the beam velocity chirp which limits the shortest bunch achievable. With a passive buncher, it may be possible to use a multi-mode dielectric tube \cite{EM2} to imprint linear velocity chirp through use of high-order wakefield modes. Furthermore, because the energy stability for keV UED can be orders of magnitude higher than MeV UED, the compressed beam may have negligibly small timing jitter with respect to the external laser, ideal for laser-pump electron-probe applications. 

This work was supported by the Major State Basic Research Development Program of China (Grants No. 2015CB859700) and by the National Natural Science Foundation of China (Grants No. 11327902, 11504232 and 11721091). One of the authors (DX) would like to thank the support of grant from the office of Science and Technology, Shanghai Municipal Government (No. 16DZ2260200).\\
*~dxiang@sjtu.edu.cn\\
$\dagger$~jzhang1@sjtu.edu.cn

\pagebreak

\end{document}